\documentclass{aa}  
\usepackage{graphicx}
\usepackage{txfonts}
\begin{document}
   \title{Pulsars with gigahertz-peaked spectra}

   \author{J. Kijak
          \inst{1},
          W. Lewandowski
          \inst{1},
          O. Maron
          \inst{1},
          Y. Gupta
          \inst{2},
          A. Jessner
          \inst{3}
          }

   \offprints{J. Kijak}

   \institute{Kepler Institute of Astronomy, University of Zielona G\'ora,
              Lubuska 2, Zielona G\'ora, PL-65-265 Poland \\
              \email{jkijak@astro.ia.uz.zgora.pl}
         \and
            National Centre for Radio Astrophysics, TIFR, Pune
              University Campus, Pune, India 
          \and
                  Max-Planck Institut f\"{u}r Radioastronomie, Auf
                 dem H\"{u}gel 69, 53121 Bonn, Germany
             }

   \date{Received ....; accepted ....}

 
  \abstract
{}
{We investigate a high frequency turn-over effect in radio spectra for pulsars with positive 
or flat spectral index.}
{Using GMRT and Effelsberg observatory, we estimated the flux density to reconstruct pulsar 
spectra.}
{We find objects that  have a maximum flux in their spectrum above 1 GHz and  
whose spectral indices are positive at lower frequencies.
Some pulsars with a turn-over in their spectrum at high frequencies
are found to exist in very interesting environments. We call these objects
gigahertz-peaked spectra pulsars.}
%
   {}

\authorrunning{J. Kijak et al.}

   \maketitle
%

\section{Introduction}
Flux density measurements and the shapes of pulsar spectra are very important in understanding pulsar 
emission mechanism. The difficulties in obtaining a true pulsar spectrum are mainly due to the 
influence of interstellar scintillation, which causes variations in the measured flux. As stated 
by Stinebring et al. (2000), pulsars are nearly constant luminosity radio sources over time spans of 
between several days and 5 years, hence interstellar scintillations must always be taken into account 
when measuring the flux.

A typical pulsar spectrum has been modelled using either a simple power law with mean spectral 
index of $-1.8\pm0.2$  or two power laws with spectral indices of -0.9 and -2.2 and a break 
frequency on average of 1.5 GHz (Maron et al.~\cite{maro00}; M00). For several pulsars 
a low frequency turn-over has been observed (Sieber~\cite{sieb73}; Malofeev et al.~\cite{malo94}).
The frequency at which this  spectrum
displays a {\it maximum flux} is called the {\it peak frequency}
$\nu_{\rm peak}$, and occurs
in the range $\sim$ 100 to 600 MHz for most pulsars with turn-over spectra.

The shape of pulsar spectrum is independant of the pulsar profile evolution. The changes in pulsar profile 
over the observing frequency do not affect the shape of the spectrum and vice versa - the observed low 
frequency turn-over is not caused by any significant profile change at the particular frequency. There is no 
correlation between the profile type and the pulsar spectrum 
(M00). However, L\"{o}hmer et al.~(2008) show that the spectrum of
PSR B0144+59 has a 'turn up' at 5 GHz to 10 GHz, which may be caused by its peculiar
high-frequency profile evolution.

Lorimer et al. (\cite{lori95}; hereafter L95) show a positive spectral index for some pulsars
between 400 MHz and 1600 MHz. By combining flux density measurements taken at frequencies above 1.4 GHz
with published data, Kijak \& Maron (\cite{km04}) 
were able to identify several pulsars who possibly had a high frequency turn-over
in the spectrum. In  Kijak, Gupta \& Krzeszowski (\cite{kgk07}), the authors presented the first
direct evidence of a turn-over in pulsar radio spectra at high frequencies using the multifrequency 
flux density measurements of these candidate pulsars. 

The spectra of the majority of known pulsars have yet to be acquired. If the pulsars with positive spectral index 
in the spectra below 1~GHz turn out to be a significant sub-group of the pulsar population, then we 
may need to re-evaluate (at least to some degree) the future low-frequency search strategies 
(for instruments such as LOFAR or SKA), otherwise we may 
overestimate the flux densities of these sources, hence overestimate the expected number of 
sources to be found in the future surveys.

In this paper, we present our observations and the information that we can derive from our pulsar radio spectra. These objects reach a {\it maximum flux} in their spectrum above 1 GHz 
and their energy decreases below 1 GHz, producing a positive
spectral index at lower frequencies.  We call these objects gigahertz-peaked spectra (GPS) pulsars.
For ten new pulsars, we also estimated the spectral index and for an additional ten  objects, we recalculated
the spectral index after considering values in the~ATNF Catalogue.


\section{Observations and results}  
Knowing that pulsars with a  turn-over around 1 GHz are relatively young (Kijak, Gupta \& 
Krzeszowski \cite{kgk07}; hereafter KGK), we planned to search for the high frequency turn-over 
effect in some recently discovered young pulsars and selected candidates from previous 
work (Kijak \& Maron \cite{km04}). Observations of these pulsars, using the GMRT and Effelsberg 
observatory helped us to study this effect. 

The observations with Giant Metrewave Radio Telescope (GMRT)
were conducted in January and February 2008 at two  
frequencies, 610 MHz and 1170 MHz, using 16 MHz bandwidth. We used the phased array mode
with 0.512 msec sampling and 256 spectral channels across the band (Gupta et al. 2000). 
To estimate the mean flux density of the pulsars,
we carried out regular calibration measurements of known continuum sources (e.g. 1822-096).
We observed nine pulsars in total intensity mode for most sources at several epochs (see Table~1). 
Some pulsars were not observed at the lower frequency 
as the expected scatter broadening was found to be comparable to or greater than the pulse 
period (see in details KGK).  Integration times of the pulsars were between 20 and 30 minutes
per source, depending on the expected flux density.

All observations at higher frequencies (2.6 GHz and 4.85 GHz, with 100 MHz and 500 MHz bandwidth, 
respectively) were made with the 100-m radio telescope of the Max-Planck Institute for 
Radioastronomy  at Effelsberg . 
We used secondary focus receivers (with cooled HEMT amplifiers) providing LHC and RHC signals
that were digitised and independently sampled every $P/1024$ s and synchronously folded 
with the topocentric pulse period $P$ (Jessner 1996).
We carried out our observation in June 2009. The typical integration time was  about 30 minutes.
To measure flux density, we carried out regular calibration measurements using an 
injected signal of a~noise diode, which  was compared to the flux density of known continuum 
sources (NGC 7027, 3C 273, 3C 286). We did not observe the sources at 8.35 GHz because of
poor weather conditions. 

None of the pulsars in our sample displays any signs of a~profile evolution at low frequencies
(see Fig.1, Fig.2, discussion and KGK), that may have caused the flux densities to be underestimated.
Interstellar scattering for all the pulsars at the lowest observing frequencies was found to be much 
less than pulse period so it shouldn't affect out results either.

The flux measurements derived
from the above observations are presented in Table~1, i.e. the flux density $F_{\nu}$, 
the error in $F_{\nu}$ and the number of measurements. The estimated 
uncertainties include  contributions from both the calibration procedure and the pulse
energy estimation.

\begin{table}
\begin{center}
\caption{Flux density measurements in GMRT and Effelsberg. The number of measurements at each frequency
are given in parentheses. ``$<$'' denotes an upper limit.}
\scriptsize
\begin{tabular}{lllcc}
\hline
PSR &  \multicolumn{1}{c}{$F_{610}$} &  \multicolumn{1}{c}{$F_{1170}$} & 
\multicolumn{1}{c}{$F_{2640}$}  & \multicolumn{1}{c}{$F_{4850}$} \\ 
 & \multicolumn{1}{c}{(mJy)} & \multicolumn{1}{c}{(mJy)}   & \multicolumn{1}{c}{(mJy)} & \multicolumn{1}{c}{(mJy)}  \\
\hline
J1740+1000            & 6.1$\pm 2.7$(4) &  0.9$\pm 0.3$(3) &  $<$0.5 &  $<$0.05   \\
B1800$-$21 &      &   & 7.60$\pm 0.50$(2) &     \\
J1806$-$2125&      &  1.5$\pm0.5$(1) & $< 0.6$ & $< 0.08$    \\
J1809$-$1917&       & 1.1$\pm 0.3$(2) & 2.30$\pm 0.30$(2) & 0.91$\pm 0.10$(2) \\
B1820$-$14 &       &       &   0.26$\pm 0.09$(1) &  \\
B1822$-$14 &  2.0$\pm 0.3$(4)$^*$  &  &  & \\  
B1823$-$13 &     &    & 3.60$\pm 0.10$(2) &   \\
J1828$-$1101&      &  &  1.20$\pm 0.20$(1) & 0.09$\pm 0.05$(2) \\
B1828$-$11 & 1.6$\pm 0.2$(2)$^*$ &  &  0.40$\pm 0.10$(2) & 0.06$\pm 0.01$(1) \\
B1832$-$06 &     & 3.5$\pm 0.6$(1)$^\star$ & 0.55$\pm 0.10$(2) &  0.09$\pm 0.05$(2) \\
J1835$-$1020           & 5.8$\pm 1.7$(1) & 2.7$\pm 0.4$(2) & 0.88$\pm 0.09$(2) & 0.14$\pm 0.08$(2) \\
B1849+00  &     & 5.5$\pm 1.5$(1)$^\star$ & 1.30$\pm 0.15$(1) &  \\
J1857+0143             &\multicolumn{1}{c}{$<0.8$} & 0.5$\pm 0.2$(2) &  0.19$\pm 0.06$(3) & 0.03$\pm 0.02$(1) \\
J1905+0616             & 2.1$\pm 0.6$(1) & 0.4$\pm 0.1$(2) &    &    \\
J1907+0918          & 1.0$\pm 0.2$(2) & 0.2$\pm 0.06$(3)& 0.04$\pm 0.02$(2) &   \\
\hline
\multicolumn{2}{l}{$*$ - GMRT 2005 and 2008} & 
\multicolumn{3}{l}{$\star$ - GMRT 2005 (KGK)}   \\
\end{tabular}
\normalsize
\end{center}
\end{table}

\begin{figure}
\unitlength=1mm
\ \put(-5,-104)
{\includegraphics{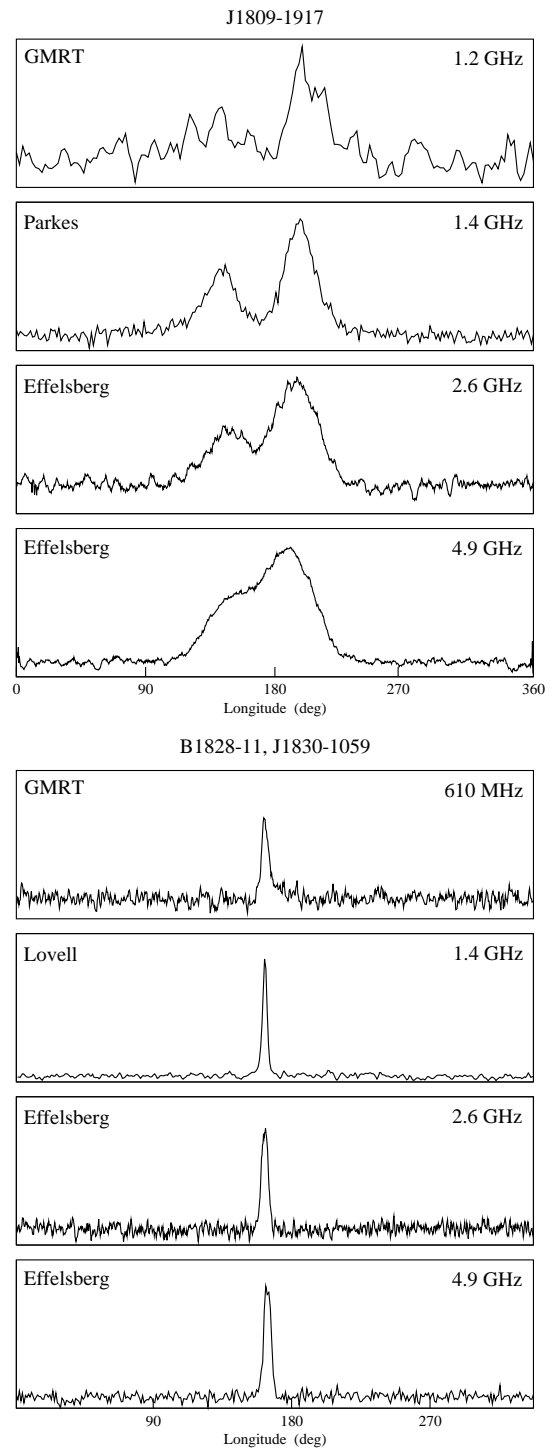}}
\ \put(-5,-201)
{\includegraphics{14274fg1b.eps}}
\vskip-8mm
\caption{Pulse profiles for PSR~J1809$-$1917 and PSR~B1828$-$11 showing no significant profile evolution with frequency (see discussion  
for details). Profiles were taken from our observations (610 MHz, 2.6, and 4.85 GHz), from EPN data base 
(1.4~GHz, see Lorimer et al. \cite{lori98} and Gould \& Lyne \cite{gould}) and from 
the CSIRO Data Access Portal (http://datanet.csiro.au/dap/).}
\end{figure}

\begin{figure}
\centering
\includegraphics[width=8cm]{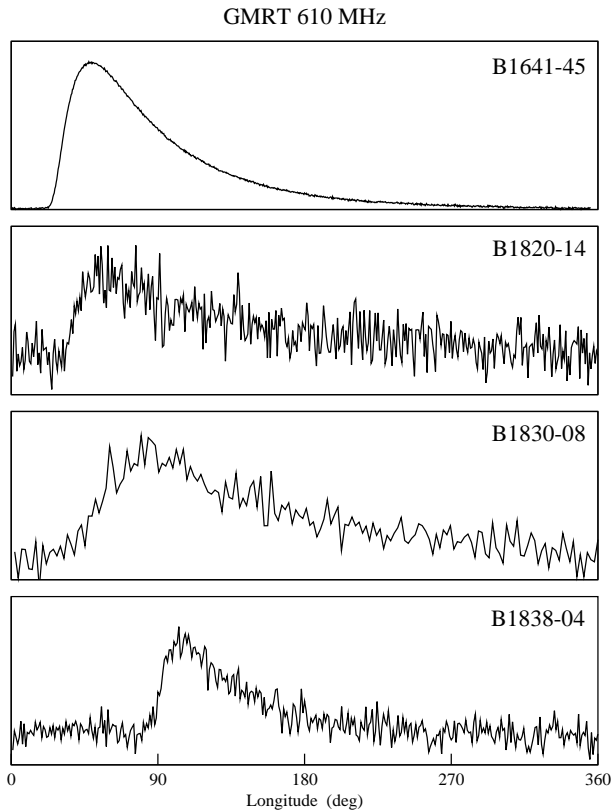}
\caption{Example profiles, from our observations at 610~MHz, showing significant pulse broadening. Three  bottom 
panels show pulsars for which our flux measurements do not agree with previously published data. For details, see 
discussion in the text.}
\end{figure}

\section{Spectral index and gigahertz-peaked spectra}

Using our results with data taken from literature
(e.g. L95; M00; McLaughlin et al. \cite{mcla02}; Hobbs et al. \cite{hobs04};
KGK), we constructed spectra for the pulsars and investigated them
by searching for those objects with {\it peak frequency} above 1 GHz. The results of our work are summarized 
in Table~2, and  a graphical representation of the pulsar spectra
can be found in Figures 3~-~5.

Table~2 presents the derived spectral indices and additional basic parameters for the  20 pulsars that 
were investigated. The second column of the table, denoted $\alpha_{\rm \small ATNF}$, contains the spectral 
indices of these sources taken from the ATNF\footnote{http://www.atnf.csiro.au/research/pulsar/psrcat/}
database (Manchester~et~al.~2005). We note that a majority of these cited spectral indices correspond to 
spectra much flatter than the usual pulsar spectra, one of the main reasons why these sources 
were chosen. However, the next column of the table ($\alpha$) indicates the spectral index of the
investigated pulsars, which was derived using our observations, along with the previously published 
flux density data. For nine pulsars, this is the~first estimate of their spectral index. For the remaining
eleven sources, our values differ significantly from those published in the ATNF database, i.e., they are
much closer to the average pulsar spectral index of $-1.8$. The main reason for this difference is 
the much wider frequency range we used (see column 4 of the table).

For pulsars with a break in the spectrum - whether a 
turn-over, or a change from flat or nearly-flat to steep spectrum
(classic broken-type spectra) - 
the spectral index $\alpha$ given in Col.3 of the table denotes the slope of the spectrum in the high 
frequency part, i.e. after the break.

For the above-mentioned cases, we decided to use a simple double-power-law model to describe the pulsar 
spectrum, and find the value of the frequency of the spectral break where applicable. Column~5 of the table, denoted
$\alpha_P$, presents the value of the spectral index of the low-frequency part of the spectrum, while the subsequent 
two columns describe the frequency range that was used to find the slope, and the approximate value of the 
frequency at which the turn-over (or break)  in the spectrum appears.

Two of the pulsars - B1054$-$62 and B1740$-$31 - were included in Table~2 only for completion reasons. 
Despite no new data have been gathered for them (all values were taken from KGK), 
we wished that the table include all the pulsars that, in our opinion, show a genuine turn-over above 600 MHz 
in their spectra.

\begin{table*}[]
\begin{center}
\caption{Spectral indices and other parameters of pulsars from a search for turn-over at high frequency 
(KGK and Table 1 in this paper). In column containing $\nu_{peak}$, the break frequency is given
in parentheses (see details in text). $\alpha$ and $\alpha_p$ were estimated using our data.}
\scriptsize
\begin{tabular}{lrlcccccccc}
\hline
PSR & \multicolumn{1}{c}{$\alpha$} & \multicolumn{1}{c}{$\alpha$} & Freq. range & $\alpha_p$ & Freq.range & $\nu_{peak}$ & 
$DM$ & Age & $\dot{E}$ & Assoc. $\&$ Remarks \\ 
    &  ATNF     &   & (GHz)       &            & (GHz)      &   (GHz)  &  (pc cm$^{-3})$ & (kyr) & (erg/s)  &    \\
\hline
\multicolumn{11}{c}{\it Normal spectra} \\
B1557$-$50 &    &  -1.4 & 0.3 - 1.4 &       &             &      & 260 &     603 & 2.8e+34 &    \\
J1828$-$1101 & & -2.5 & 1.4 - 4.9 & &  &    & 607 & 77 & 1.6e+36 & \\
J1835$-$1020&     & -1.6 & 0.6 - 4.9 &     &            &      &  113 &     810 & 8.4e+33 &     \\
\multicolumn{11}{c}{\it Flat-normal or unclear spectra} \\
B1641$-$45 &    &  -0.9 & 0.6 - 1.4 &       &             &      & 478 &     359 & 8.4e+33 &    \\
J1740+1000 & 0.9  & -2.9 & 0.4 - 1.4 &   &   &     &  24 &     114 & 2.3e+35 &   see Fig.4   \\
B1800$-$21& -0.1 & -1.1 & 1.4 - 11  & 0.0  &  0.6 - 1.4 & (1.6)& 234 &     15.8 & 2.2e+36 & HESS, X:PWN, SNR    \\
B1828$-$11 & -0.5 & -2.4 & 1.4 - 4.9 & 0.5 & 0.6 - 1.0  & 1.2  &  161 &     107 & 3.6e+34 &     \\
B1820$-$14  & -0.8& -2.5 &  0.6 - 2.6 &     &           &      &  651 &    3750 & 3.6e+33 &  corrected L95(610MHz)    \\
B1830$-$08 & -0.4 & -1.8 & 0.6 - 4.9 &     &            &      &  411 &     147 & 5.8e+35 & corrected L95(610MHz)  \\  
B1832$-$06 & -0.4 & -1.9 & 1.0 - 4.9 &     &            &      &  473 &     120 & 5.6e+34 &    \\
B1838$-$04&  -0.4 & -1.7 & 0.6 - 8.4 &     &            &      &  325 &     461 & 3.9e+34 & corrected L95(610MHz)   \\
B1849+00  &       & -2.4 & 1.4 - 4.9 &     &            & 1.3? &  787 &     356 & 3.7e+32 &    \\
J1857+0143&       & -1.5 & 1.0 - 2.6 &     &            & (1.2) &  249 &      71 & 4.5e+35 & 3EG, HESS \\
J1905+0616&       & -1.9 & 0.6 - 1.4 &     &            & 0.6? &  256 &     116 & 5.5e+33 &    \\
J1907+0918& -0.3& -2.2 & 0.6 - 2.6 &     &            & 0.6? &  358 &      38 & 3.2e+35 &    \\
\multicolumn{11}{c}{\it Turn-over and GPS from {\rm KGK}} \\
B1054$-$62 &    &  -3.2 & 0.9 - 1.4 & 1.0  &  0.3 - 1.0 & 0.95 & 320 &    1870 & 1.9e+33 & HII \\
B1740$-$31 & -0.9& -2.0 & 0.6 - 1.4 & 0.6  &  0.3 - 0.6  & 0.6  & 193 &     317 & 3.4e+32 &     \\
B1822$-$14 & -1.4 & -0.7 & 1.4 - 4.9 & 0.6  & 0.6 - 1.4 & 1.4  &  357 &     195 & 4.1e+34 &3EG, HESS \\
B1823$-$13 &      & -0.9 & 1.6 - 11  & 1.1 & 1.0  - 1.6 & 1.6  &  231 &      21 & 2.8e+36 &3EG, HESS, X:PWN \\ 
\multicolumn{11}{c}{\it New GPS} \\
J1809$-$1917 &    &  -1.9 & 2.6 - 4.9 & 4.3 & 1.1 - 1.4 & 1.7  &  197 &     51.3 & 1.8e+36 &  X:PWN,HESS  \\    
\hline
\end{tabular}
\normalsize
\end{center}
\end{table*}

\begin{figure*}
\centering
\caption{The spectra of pulsars from Table~2. Empty circles  - L95, M00 and K98,
full circles - KGK and this paper, diamonds  - ATNF, and cross is the result for B1832$-$06 from L95, 
which we consider to be doubtful because no profile for this measurement was published (see text for details)}
\vskip5mm
\includegraphics[width=16cm]{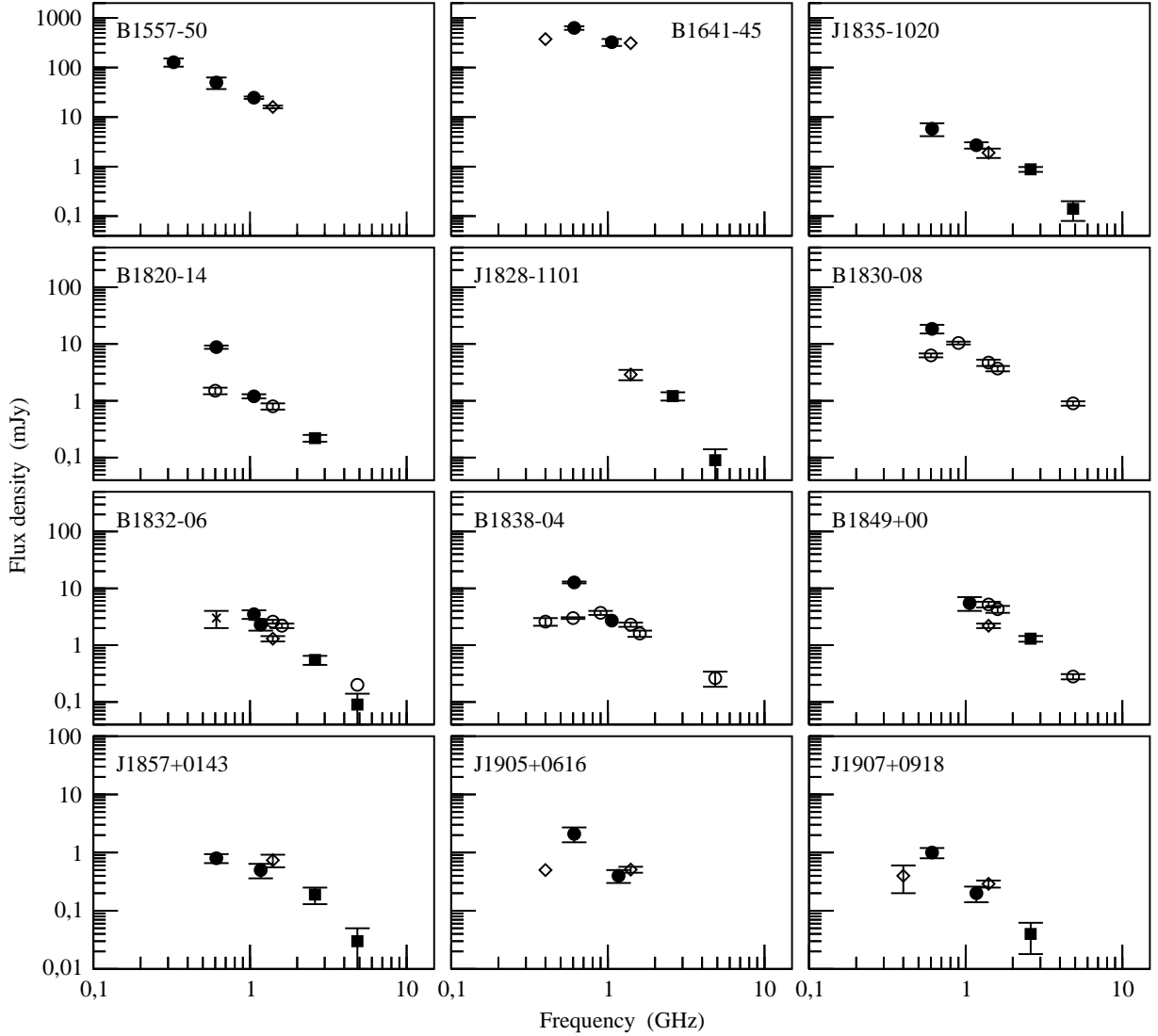}
\end{figure*}

The spectra of the remaining 18 sources can be found in Figures~3~to~5. In these plots, full squares
represent our Effelsberg observations, full circles our GMRT data, empty circles are measurements 
from L95 and M00, diamonds indicate  ATNF data (see references therein), and the star symbols
are the measurements of Hobbs~et~al.(2004).

For clarity we distingushed four groups of pulsars in Table~2:

{\bf Group one} consists of 
three sources that are regular steep spectra pulsars (see Fig~3), which were included in our 
program because they were young and moderate- to high-DM objects that previously only had flux 
measured at single frequency (1.4~GHz).

{\bf The second and largest group} in Table~2 consists of sources that were believed to have 
either flat spectra or a~positive spectral index based on previously available data (L95 and M00). 
The majority of the spectra for this group can be found in Fig~3. In most cases, adding our 
observations, especially at the higher frequencies, 
changed the appearance of spectra to regular steep spectra, such that a negative spectral index 
occured close to the cannonical $-1.8$; we believe that this occured because the spectra of these 
pulsars were previously known only over relatively narrow frequency ranges. 

In a few cases (the pulsars that have the values of $\nu_{peak}$ cited), the shape of the pulsar spectrum, 
especially at lower frequencies, remains unclear, while at frequencies higher than  $\nu_{peak}$ 
the spectra behave like those of regular pulsars. Some of these objects were previously believed to have flat 
spectra because the only flux measurements available were at frequencies close to the spectral break frequency. 
For these sources, the existence of the turn-over (or break) in the spectrum may be - at the moment -  is
at best doubtful.

In three cases, B1820$-$14, B1830$-$08, and B1838$-$04, our measurements at 610~MHz
differ significantly from those previously published in the literature, i.e., Lorimer~et~al.~(1995).
We compared the profiles of these pulsars, which can be found in the EPN 
Database\footnote{http://www.mpifr-bonn.mpg.de/div/pulsar/data/browser.html} (Lorimer et al. \cite{lori98}),
with our profiles acquired 
at GMRT (that can be found on Fig~2)\footnote{full sample of the acquired profiles will be published 
in a~forthcoming paper considering the scattering phenomenon itself; Lewandowski~et~al. in-prep}. 
Our belief is that the flux density measurements published earlier are doubtful, and  probably 
underestimated because of problems finding the proper 
baseline of the profile. The same applies to the L95 flux measurement of B1838$-$04 
at 410~MHz. We note that the authors of that paper were aware of the problem, and note their 
measurements as heavily affected by scattering.

Similar results are found for  PSR~B1832$-$06, and the observation at 610~MHz - measurement marked as a cross in the 
appropriate panel of Fig.3. The value of the flux density at this frequency was published by L95, but the 
corresponding profile is in neither the paper nor the EPN Database. Judging by the relatively broad profiles 
acquired at 1.4~GHz and 1.6~GHz, and taking into consideration this pulsar's DM value of 463~pc~cm$^{-3}$, we 
doubt whether the published measurement is a reliable value, as it probably suffers from the effects 
described above.

In our fits of the spectral slopes for the purposes of Table~2, we decided to omit these values, 
and use our measurements instead where applicable.

Three special cases thet we wish to highlight in this group are J1740+100, B1800$-$21, and B1828$-$11.

\vskip3mm

{\it Pulsar J1740+1000} was first observed by McLaughlin~et~al. (2002) and reported to have an unusually high positive
spectral index of +0.9. During our observations, we were however able to successfully observe this source at 
GMRT at two frequencies, 610 and 1100~MHz, and found that our values sharply contradict the published findings 
(see Fig~4), i.e. our value measured at 1100~MHz is lower than the previously reported value at 1.4~GHz by a factor 
of ten. We tried to follow this using the Effelsberg radiotelescope at high frequencies (2.6~GHz and 5~GHz), 
but we were unable to detect the pulsar. Triangles on the plot denote upper limits derived from the 
gathered data.

The only explanation we have for this discrepancy is that McLaughlin~et~al.(2002) reported during 
their observations that the pulsar was undergoing strong scintilations, which we did not see in our data.
These scintillations apparently amplified the pulsar's signal during their 1.4~GHz by a huge factor, which 
in turn changed the appearance of the spectra of this - based on our data - otherwise average pulsar. 
Nevertheless, this case requires further study, as we do not have an explanation of why these strong 
scintillations were present during the initial observations, and there was no evidence whatsoever during 
our project.

\vskip3mm

{\it PSR~B1800$-$21}, whose spectrum can also be found in Fig~4, is believed to be a young, Vela-like pulsar, 
associated with a supernova remnant, because it displays clear indications of a pulsar wind nebula in X-ray 
observations (Kargaltsev~et~al.~2007). The radio spectrum of this source displays a clear break around
1.6~GHz. The spectrum appears to be flat at low frequencies, especially when one excludes the flux measurement
at 408~MHz (denoted by a cross on the plot) which we consider at best doubtful. The inspection of 
the profile for this observation in the EPN Database clearly shows that at such a low frequency this pulsar
undergoes a significant scatter-broadening, and we doubt that a proper profile baseline could be found in 
these data, hence the flux measurement may be scrambled.

The spectral index above the break for B1800$-$21 is $-1.1$, which is lower than average but not alarmingly so.

\vskip3mm

{\it B1828$-$11} is a pulsar that could be considered to display a high frequency 
turn-over in the spectrum, where the {\it peak frequency} is around 1.2~GHz (as suggested in Fig.~4), 
but the spread of measurements around this frequency causes that this interpretation is questionable. One would need 
a reliable measurement at frequencies lower than 600~MHz to decide between a broken type spectrum and a 
turn-over spectrum. This measurement should be possible, as our profile at 610~MHz suggests that the pulsar is not 
undergoing significant scattering  - DM is only 161~pc/cm$^3$, one of the smallest in our sample, and the measured 
scattering at 610~MHz is $\tau_{sc}\sim 5$~ms for our GMRT data (we used the method from L\"{o}hmer et al. 2004
with their $PBF_1$ function to estimate this value), slightly above 1\% for a pulsar with 405~ms period 
(see also Fig.~1). Despite this, except for an upper limit from L95 there is no measurement as of yet,
a situation that we plan to change in the near future.

Until then, we decided to classify this pulsar as an unclear case, probably a broken-spectra type similar 
to B1800$-$21.

\vskip3mm

{\bf The third group} in Table~2 are previously known pulsars with a high frequency turn-over 
in the spectrum (from KGK). On the basis of these new observations, we corrected the values of 
$\nu_{peak}$ and both spectral indices (for the positive and negative part of the spectrum) for pulsars 
B1822$-$14 and B1823$-$13 (see Fig.~5). 

We note that our fit for the positive spectral index of B1823$-13$ did not include the L95 
measurement at 610~MHz, since no profile has been published (marked by a cross; see above-mentioned cases of 
B1832$-$06 and B1800$-$21). Its agreement with the fit is purely coincidental.

{\bf The last pulsar} in the table is found to display gigahertz-peaked spectra
for the first time. Flux had previously been measured for PSR~J1809$-$1917 at only one frequency; 
new data (see Fig.~5)
clearly indicate a peak in the spectrum around 1.7~GHz. The profile of this pulsar shows evidence of neither 
significant scattering nor dispersion smearing. Dispersion at 1170~MHz for a DM=197 pc/cm$^3$ pulsar is
1~ms per MHz. For our GMRT data at 1.17~GHz, the scattering comes out as $\tau_{sc} \sim 3$~ms, 
similar to what can be measured for a profile from the ATNF database at 1.4~GHz (2~ms, conf. Fig.~1). 

Another source of measurement error may be the interstellar scintillation (ISS). For PSR~J1809$-$1917,
we estimated the diffractive timescale at 1170~MHz (Lorimer \& Kramer, 2005; see also Section \ref{scint} 
for details) to be of the order of 100 seconds, which means that any flux density variations caused by 
DISS will be averaged-out in a 30-minute integration. At the same time, the refractive timescale 
is found to be $\sim 2.5$~day, and the expected refractive modulation index is relatively small,
$m_{RISS} \simeq 0.28$ (all these are of course crude estimations). 
Since our flux density measurement is an average of two observations separated by a few days, 
the flux density at 1170~MHz that we obtained should be reliable.

As for our higher frequency observations at 2.6 and 4.8~GHz, the DISS timescale can be estimated to be 
260 and 1300 seconds, respectively, while for a refractive ISS we obtained 0.5 and 0.25 days. 
In this case, both RISS and DISS may influence the measured values. We observed the pulsar at each frequency 
for a total of 1 hour, which we hope is long enough to alleviate the effects of diffractive scintillations, 
as the observation time is 
several times longer (in the case of 2.6~GHz observations) down to at least three times longer (4.8~GHz) 
than the DISS timescale. On the other hand, RISS may be more significant at these frequencies, especially 
since the expected modulation index is larger than at 1170~MHz (0.43 at 2.6~GHz to 0.62 at 4.8~GHz), although
we hope that even if ISS would cause errors in our measurements it would not be to a degree large enough to
change the appearance of the spectrum in any significant way.

\vskip3mm

For our entire sample, the average spectral index of the pulsars with typical spectrum 
shown in Table~2 is $-$1.8, i.e. exactly equal to the global average value for all the pulsars 
with known spectra. For the pulsars with a turn-over 
in their spectrum the spectral index at lower frequencies varies from +0.5 to +4.3 (with exclusion 
of PSR~B1800$-$21), and the value of $\nu_{peak}$  ranges from 600~MHz to 1.7~GHz.

\begin{figure}
\unitlength=1mm
\ \put(7,350)
{\includegraphics{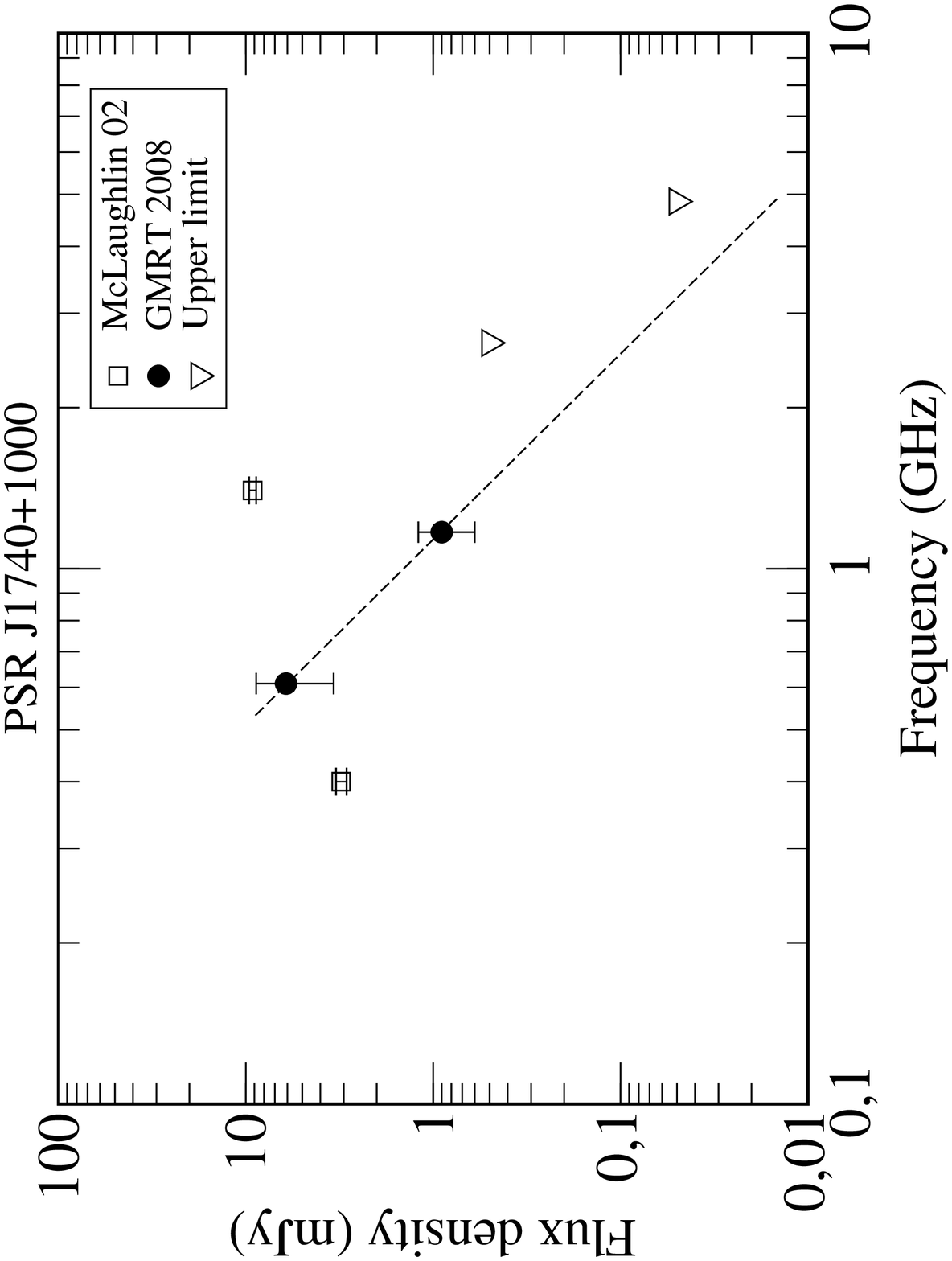}}
\ \put(7,290)
{\includegraphics{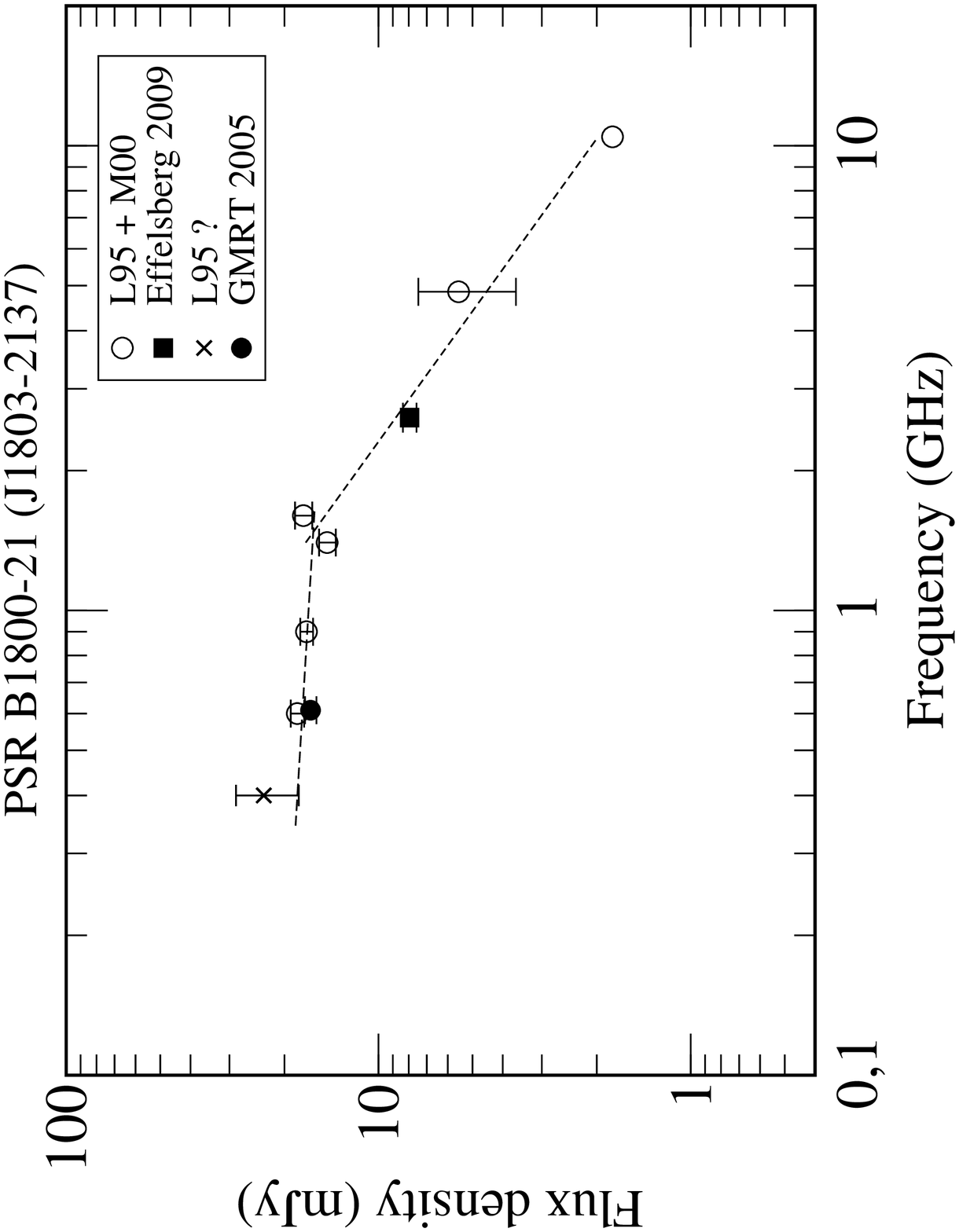}}
\ \put(7,230)
{\includegraphics{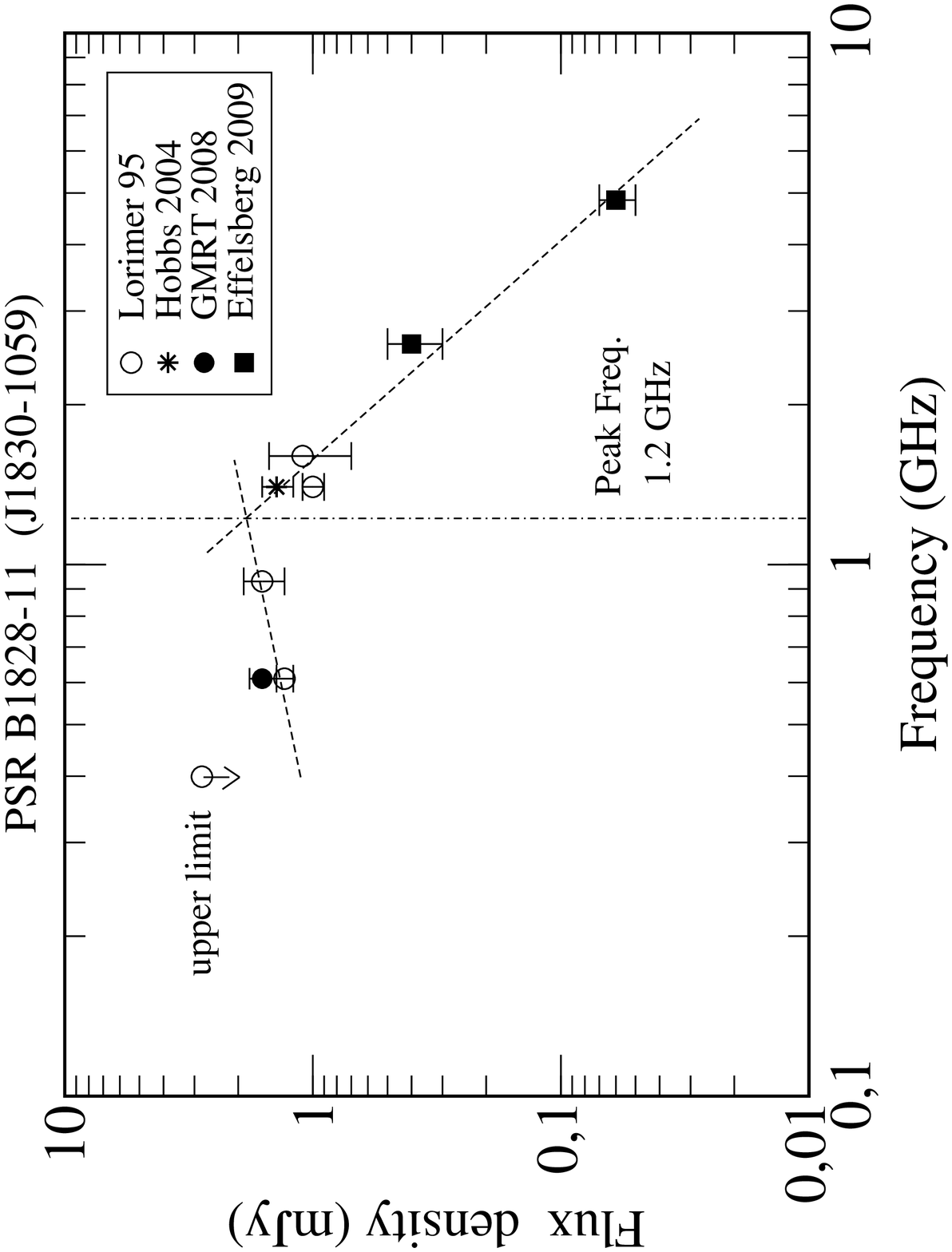}}
\vskip-170mm
\caption{Spectra of PSR~J1740+1000, PSR~B1800$-$21 and PSR~B1828$-$11. Cross on the 
spectrum of B1800$-$21 represents a flux measurement from L95, for which no profile 
has been published, and we consider it doubtful (see discusion and similar case in Fig~3).}
\end{figure}

\section{Discussion}

It remains unclear whether the cause of
the turn-over is  some kind of absorption in the magnetosphere, efficiency loss of 
the~emission mechanism, or an interstellar effect (Sieber \cite{sieb73}). 

To confirm or disregard one of the above possibilities, we need to construct the spectra for a
large number of pulsars, especially at frequencies below 1 GHz where positive spectral index may
correspond to a possible turn-over at high frequency. This and some connection with high frequency flux 
measurements will allow us to estimate the {\it peak frequency} and the~shape of the spectrum, which 
may be helpful when we attempt to identify the origins of the spectral turn-overs.

There are also many other reasons that for an apparent decrease in the flux density at lower frequencies.

\subsection{Interstellar scattering and corrected spectra}

The scattering phenomenon causes pulse profiles to become broader, i.e. pulses to attain roughly exponentially
decaying scattering tail. It has been shown that the characteristic broadening of the pulse, $\tau_{sc}$, 
depends on both the observing frequency, as well as the disperssion measure, the empirical relation given by 
Bhat~et~al.~(2004) being $\log \tau_{sc} = -6.46 +1.054\log (DM) + 1.07 \log^2(DM) -3.86 \log \nu$. For high DM pulsars 
at low frequencies, $\tau_{sc}$ may become so long that one will not see any pulsed emission, when the scattering time
is greater than the pulsar period by a significant factor.

The most difficult analyses of pulsar spectra occur when the scattering time is close 
to the pulsar period. The classical method of flux density measurement requires that the baseline of the profile 
is found to subtract the background flux. One has to remember that $\tau_{sc}$ is the characteristic timescale of 
the exponential scattering tail decay, the timescale across which the~tail flux density decreases by 
a factor of $1/e$.
If $P$ is a pulsar period and we assume that $\tau_{sc}=0.5 P$, the tail will still contribute $1/e^2$ 
($\sim 0.135$) of the peak flux after a full pulsar period, i.e. to the next pulse, or to the main pulse in the
integrated profile case. This means there can be no proper baseline of the profile found.

In observations with a high noise level where it is difficult to estimate the value of 
$\tau_{sc}$, one can clearly see  pulsed emission, but at the same time misinterpret a noisy, low-slope 
scattering tail as a baseline level. This can lead to an underestimate of the pulsar flux, and hence a change in 
the appearance of a pulsar spectrum. The significance of this effect would increase for weaker pulsars  
(more noise in the profile), and the~value of $\tau_{sc}$ increased. For a given pulsar, this
would imply that as the~frequency decreases, the flux becomes more underestimated.
In~cases such as this, the only way of measuring the~pulsar flux is by means of continuum (interferometric) methods
(see for example Kouwenhoven, 2000).

For our pulsars, we are convinced that this effect if present, has a minimal impact on the flux measurements 
(see Figs.~1 and 2 for some examples, and the discussion of a case of J1809$-$1917 in a previous 
section), hence it would not change the appearance of the spectra in any 
significant way.

In the case of some old observations, such as L95 or M00 (see the discussion on pulsars B1820$-$14, B1830$-$08, 
B1832$-$06, and B1838$-$04 in the previous section), one has to approach the results carefully,  
always analyse the pulse profile before making any assumptions, and carefully study the previous analysis approaches,
because in some cases authors realised the measurements may be heavily impacted by the scattering effect, 
and noted this accordingly.

\begin{figure}
\unitlength=1mm
\ \put(7,350)
{\includegraphics{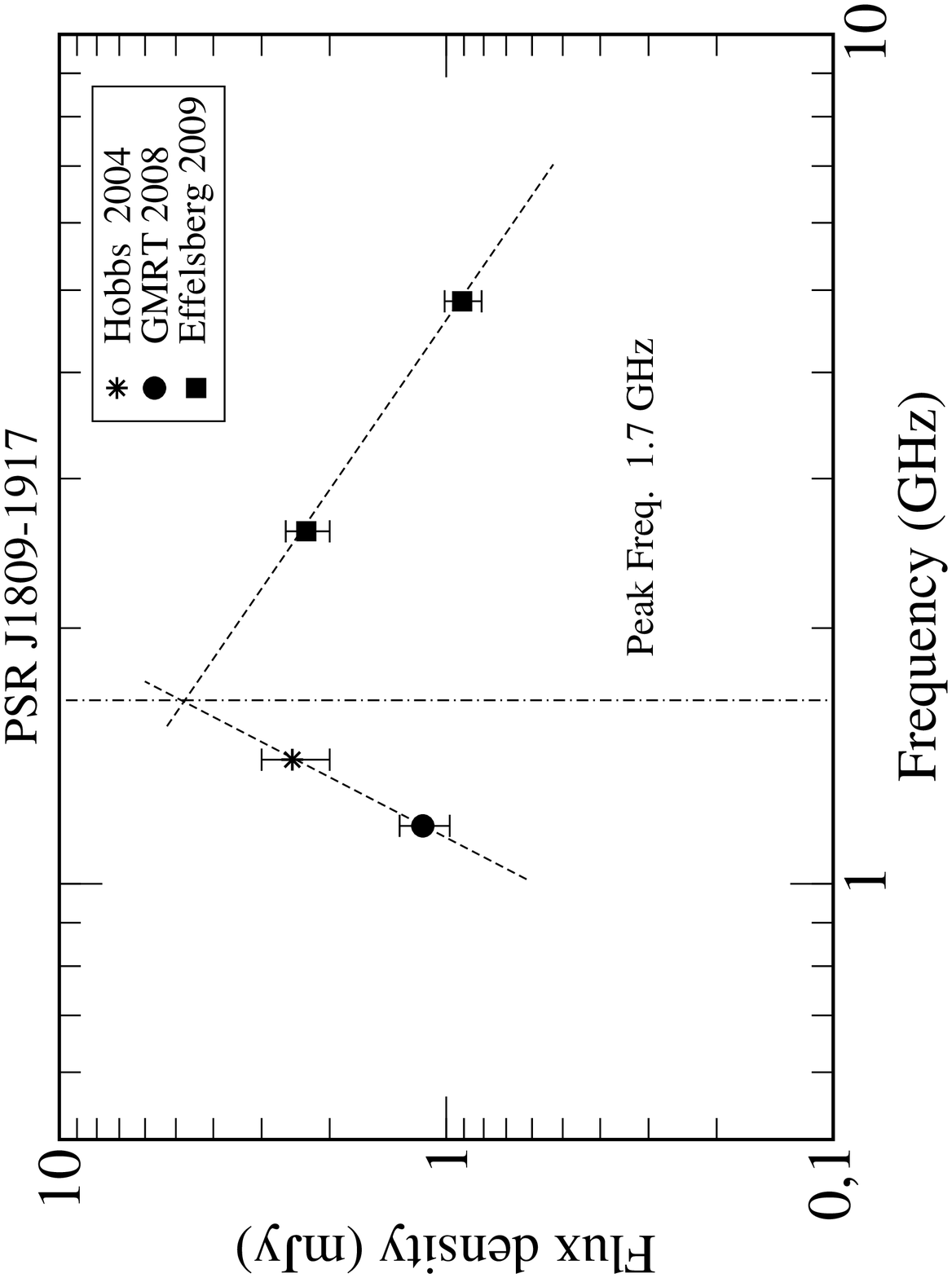}}
\ \put(7,290)
{\includegraphics{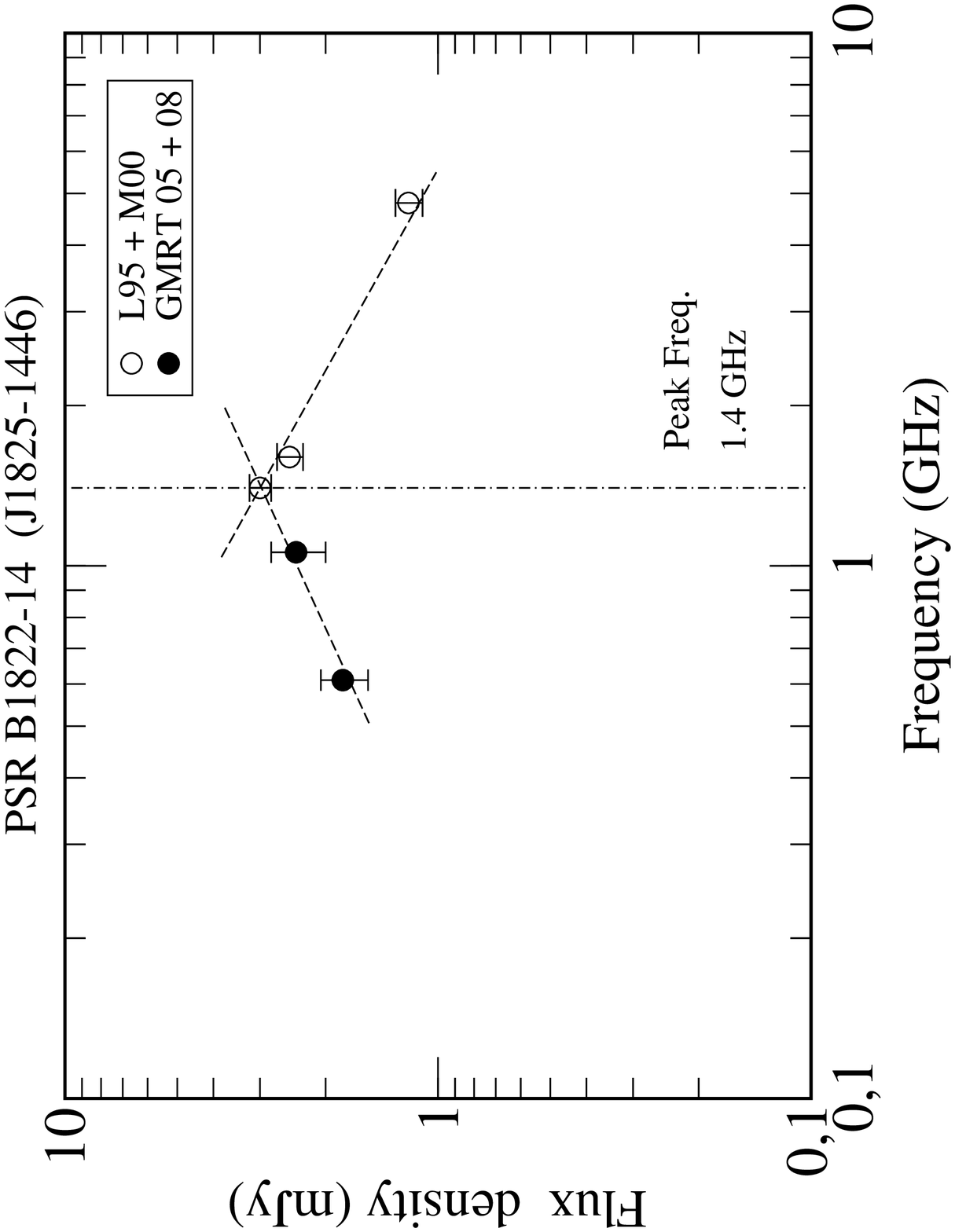}}
\ \put(7,230)
{\includegraphics{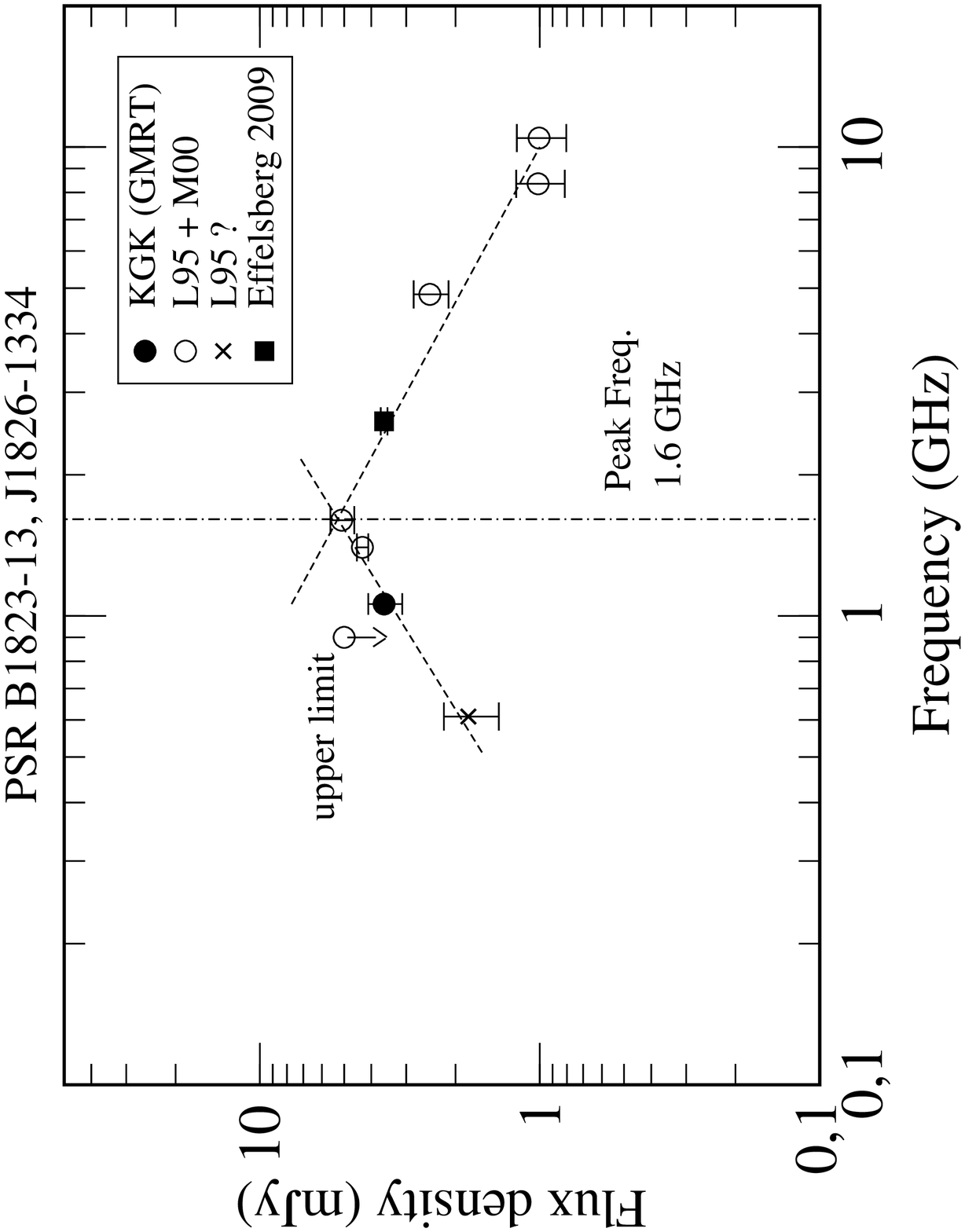}}
\vskip-170mm
\caption{Pulsars with the gigahertz-peaked spectra. Spectra of B1822$-$14 and B1823$-$13 are updated with 
respect to those published in KGK. For J1809$-$1917, this is the first presentation
of its GPS-like spectrum.}
\end{figure}

\subsection{Scintilation and decreasing/increasing energy in spectrum \label{scint}}

Interstellar scintillations may either amplify or suppress the~pulsar signal during the observation, leading to 
an incorrect measurement of the pulsar flux density. The only way to take this effect into account is to repeat
the observations to derive an average flux value. This is especially important for pulsars that display a strong 
flux modulation.

The majority of our pulsars are objects with relatively high to very high dispersion measures (see Table~2), which 
would indicate that their transition frequency (frequency of the  switch between strong and weak scintillation
regimes, see Lorimer~\&~Kramer,~2005, for a review) is very high, well outside the frequency range the radio 
pulsars 
are usually observed. This implies that they are in the strong scintillation regime, and show both diffractive 
(DISS) and
refractive scintillations (RISS). For refractive scintillations at the observed frequencies, one would expect 
very low 
modulation of the flux density as we observe at frequencies much lower than the transition frequency. 
To estimate the scintillation parameters, we used a~few formulae from Lorimer \& Kramer 
(2005). For a~given pulsar, one has to know its distance, which can be found in the ATNF
database. To estimate scintillation parameters at a~certain frequency, one can start by estimating 
the decorrelation bandwidth (equation 4.40 in Lorimer \& Kramer, 2005; repeated after Cordes et al., 1985), 
then assuming an average scintillation speed of 100~km/s  one can estimate the DISS 
timescale (eq. 4.49; Gupta et~al., 1994). Knowing the decorrelation bandwidth, one can estimate the 
scintillation strength parameter $u$ and the value of the RISS timescale (eq. 4.46), as well as the 
refractive modulation index (eq. 4.47). These will of course be very crude estimate. In contrast any kind 
of scintillation parameters derived from the analysis of the real data would be much more accurate, 
although usually no such information is available, for very weak sources these observations
being very difficult to perform.

To illustrate the estimation process we take an hypothetical average pulsar from our sample and assume that 
DM equals 350~pc/cm$^3$, which can be used to estimate the pulsar distance from the Taylor \& Cordes (1993) 
galactic electron density model (value available in ATNF database).  In our example, we assume that 
$d=7$~kpc, which should be a typical value. Using the method described above one can then estimate that 
the DISS timescale varies from 20 seconds at 400~MHz, to 90 seconds at 1.4~GHz, and 370 seconds at 4.8~GHz. 
At the same time, the RISS timescale drops from $\sim 30$ days at 400 MHz to 2.5 days at 1.4~GHz to 
3.5 hours at 4.8~GHz, and the~RISS modulation index rises slowly with frequency, from 0.14 at 400~MHz 
to 0.56 at 4.8~GHz.

As one can see for this pulsar, the diffractive scintillations should cause very little trouble, even at the 
highest frequencies, assuming that the integration time is of the order of at least half an hour (1800 minutes, 
which is what we used for most of our observations). The RISS may cause problems at high frequency, where 
their timescale drops to an order of a~few hours, and the modulation rises with frequency. The only way 
to overcome that issue is to repeat the observations and average a larger sample of measurements. Owing to 
telescope time limitations we decided to perform only two observations in most cases; this should give us a reliable
estimate of the pulsar flux, especially since for the purposes of the spectra construction and finding
the turn-overs, the high-frequency end of the spectrum rarely causes any problems.

The only pulsar in our sample with a low DM is PSR~J1740+1000, which was included because of its
positive spectral index (McLaughlin et al.~\cite{mcla02}). Interstellar scintillation apparently 
for this source has a significant effect on the measured flux values and the appearance of 
its spectrum (see discussion in Section~3). This shows how ISS can influence the~shape of the spectrum in 
some very unlikely - but not impossible cases. 

If ISS were not taken into account and, for example, during the lower frequency observations 
one observed the pulsar around its scintillation minimum (i.e. maximum suppression of the pulsar signal), while, 
coincidently, higher frequency observations were conducted during ISS maximum (i.e. the maximum amplification of
the signal), an artificially positive spectral index might be observed. 

The above example would of course be the most extreme event, but the more troublesome are the intermediate cases, 
where scintillations would cause a change of spectral slope in some frequency range, changing the shape of the 
spectrum in not so obvious, but significant ways. This can be a potential source of errors for pulsars 
with very few measurements, especially when the number of frequencies at which they were observed is small.

As the latter applies to the majority of c.a. 300 pulsars for which spectra have been constructed, one has to approach 
both the published, as well as new (own) observational results carefully and make every effort to exclude 
the influence of ISS from the data.

\subsection{Pulsar spectra - the shapes}

Only a handful of the papers published in the past 40 years, have studied the 
characteristics of the pulsar radio spectra. This applies to both the observational results, as well as 
theoretical works. As mentioned in the Introduction, pulsar spectrum is usually described in terms 
of a simple power-law, with low frequency turn-over or the presumed cut-off at extremely low frequencies. 
There are also several known cases of broken-type spectra, which resemble the simple power-law, but 
the spectrum becomes steeper at higher frequencies.

The first comprehensive study of the shapes of pulsar radio spectra was performed by Sieber (1973). In addition
to interpreting the very low frequency cut-off, which may be caused by a loss of coherence below a critical 
frequency, he proposed two mechanisms that may produce the low frequency turn-overs in the spectra. 
One of them is synchrotron self-absorption in the 
magnetosphere and the other is thermal absorption. He tried to explain the pulsar spectra (in the 
frequency range from roughly 80~MHz to 8~GHz) using both theoretical models, along with purely observational 
simple or two power-law models. In this frequency range, he found cases of pulsars with maximum energy 
(turn-over) at frequencies below 600~MHz, and he modelled their spectra using either synchrotron self-absorption 
model or thermal absorption. Nevertheless, in his sample (of 27 pulsars), the spectra of the majority of pulsars 
can be described by a simple power-law, which can be explained by the lack of information at low enough frequencies.

The pulsar with the highest peak frequency in his sample was the Vela pulsar, which showed the maximum energy 
around 600~MHz, and the best fit for the spectrum was the thermal absorption model. 

To our knowledge very few additional analyses have been performed since Sieber (1973), from both theoretical 
and observational point of views, to explain or describe the turn-over phenomenon.

Malofeev~(1994) presented the spectra of 45 pulsars (some of which had been considered earlier by Sieber, 1973), 
and found that for c.a. ten more pulsars there is maximum energy in the spectrum, which he did not however 
attempt to explain using any model capable of describing the turn-over effect.

From the theoretical point of view, low frequency turn-overs (and other features) in the pulsar 
spectra were addressed by Petrova (2002, 2008), who pointed a few phenomena in the~pulsar magnetosphere 
that could be responsible.

As for the external effects that could be responsible for the dampening of pulsar radiation, one has to mention the
work of Kechinashvili~et~al.~(2000). These authors tried to describe the eclipses of the binary pulsar B1957+20, 
where the pulsar radiation is absorbed by the pulsar-wind powered magnetosphere of the companion star. 
While they considered only a special case of a binary pulsar, one has to note that similar effects can 
occur in any environment, with magnetic fields, that is close enough to the pulsar to be powered by its wind.

Finally, we note that L\"{o}hmer~et~al.~(2008) considered a~heuristic model 
of pulsar radiation, consisting of a superposition of a large number of short pulses of only nano-second duration.
This seems to describe the majority of observed pulsar spectra quite naturaly, but does not work for 
pulsars with turn-over effects.

\subsection{GPS and X-ray or high energy observations}

Some pulsars with turn-overs in their spectrum at high frequencies have been shown to reside in  very interesting 
environments. PSR~B1054-62 lies behind or within a dense HII region (Koribalski et al., \cite{kori95}). 
The environment of PSR~B1823$-13$ appears to have peculiar properties in both radio observations 
(Gaensler et al., \cite{gaen03}), as well as X-ray data (Pavlov et al. \cite{pav08}, Kargaltsev~et.~al., 2007), 
where the results may indicate the existence of a compact pulsar wind nebula (PWN). For another Vela-like 
pulsar PSR~B1800$-$21, a case of a broken (flat-normal) spectrum according to our data, we also have a clear 
indication of a PWN surrounding it. The same is true for the new GPS pulsar J1809$-$1917 
(Kargaltsev \& Pavlov 2007, see also Smith~et~al.~\cite{smi08} and references therein), 
although this source appears to be a bit older than those mentioned previously, it may still be considered 
a young pulsar (with characteristic age of $\sim 500$~kyr).

At this point, we note that the Vela pulsar itself has a peak frequency at a relatively high 
value of $\sim 600$~MHz (Sieber, 1973).

For a few other pulsars included in our project (with at least one GPS pulsar amongst them), 
it seems that they are  coincident with known but unidentified X-ray sources from the 3rd EGRET Catalogue, 
or HESS observations (see remarks in Table~2, associations also taken from Smith et~al.~\cite{smi08}).
This may indicate that the turn-over phenomenon is associated with the enviromental conditions
around the neutron stars rather than being related intrinsically to the radio emission mechanism. Whether
the phenomenon is caused by the thermal absorption proposed by Sieber (1973; see the previous subsection), 
or the cyclotron resonance damping (Kechinashvili~et~al.~2000) remains to be seen. It 
is also possible that the Gigahertz-Peaked Spectra and pulsars that were previously assumed to 
have {\it broken spectra} (flat or close to flat low frequency part and steep high frequency spectrum), 
may be only different manifestations of the same phenomenon.

\section{Conclusions}

We have shown that pulsars with high frequency turn-overs exist, which we have referred to as
gigahertz-peaked spectra (GPS) pulsars. To a few cases published earlier by KGK, we have added 
a new pulsar (PSR~J1809-1917) which increases the~number of GPS pulsars to five.
During the course of our project, we have either constructed or corrected the spectra of several other 
pulsars; some of these had been previously assumed to have  flat spectra, whereas our observations indicate 
that they have either a standard steep spectrum or a broken spectrum.

A significant number of GPS and broken-type spectra pulsars appear to have interesting environments
that imply that the~main cause of turn-overs may be absorption occurring in the~close proximity 
of the pulsar. This would make those objects similar to other known GPS compact radio sources.

KGK suggested that the {\it peak frequency} of pulsars with turn-overs is correlated with both pulsar
age and DM. On the~basis of the available data and apparent association of the majority of these pulsars with
peculiar environments, we propose that the correlation with DM is purely selection effect. Turn-overs are rare
phenomena, hence (since a larger DM usually corresponds to a larger distance) we search a larger area of our 
Galaxy, finding more and more of these cases the larger the~DM~is.

The correlation of the peak frequency with the pulsar age, also proposed by KGK, 
may be caused by the 
fact that younger pulsars are more likely to reside in a dense environment that is providing 
some kind of absorption, leading to apparent turn-overs in the spectra.
On the other hand, this does not 
mean that all of the young pulsars have high frequency turn-over (GPS), since detailed studies of pulsar 
environments show that they usually have highly asymmetric geometry (see for example Pavlov \& Kargaltsev, 2008).
Taking this into account, it is easy to understand that even for a single pulsar, depending on the direction of
the line-of-sight, pulsar radiation will, or will not, undergo absorption effects leading to GPS-type 
spectrum. Consequently, when looking at the  whole young pulsar population, only some of them (i.e. 
pulsars for which the line-of-sight crosses strongly absorbing regions of their asymmetric environments) 
will be found to have high frequency turn-over.

Therefore for the general pulsar population we do not expect to find any correlation between 
pulsar age and the {\it peak frequency} (or other turn-over parameters, such as the spectral index), although
for pulsars with high frequency turn-overs one can expect the age to affect these 
parameters - this may be due to the motion of the pulsar, or the evolution of the pulsar itself, as the latter
will affect the way in which the pulsar interacts with its surroundings, and hence will change its properties.

Summarizing, we believe that  GPS pulsars are very interesting targets of
study for a large portion of the pulsar  astronomer community, both
because of their peculiar  environments, and the effect the existence of
such sources may have on future activities (pulsar search surveys;
mentioned in the~Introduction).

Further investigation is clearly required and the combination of continuum (radio and X-ray) 
observations and pulsed flux density measurements for these objects may be the clue to understanding 
this phenomenon. Such a combined analysis might establish that both pulsars with GPS spectra may be interesting 
candidates for continuum (especially X-ray) studies, and pulsars with interesting environments may 
have remarkable spectrum shapes (in cases where the shapes are unknown or not well-studied over 
a wide range of frequecies).

Owing to the significance of the scattering phenomenon at lower frequencies and the pulse broadening 
that it causes, 
the~flux densities of several of the aforementioned pulsars simply cannot be measured by the usual 
means with appropriate accuracy. The only way to tackle this problem would be to conduct continuum 
interferometric observations  (in a similar way to Kouwenhoven, 2000), as we intend to do so in the near future.

\begin{acknowledgements}
We thank the staff of the GMRT who have made these observations
possible. The GMRT is run by the National Centre for Radio Astrophysics
of the Tata Institute of Fundamental Research. 
Based on observations with the 100-m telescope of the MPIfR 
(Max-Planck-Institut f\"{u}r Radioastronomie) at Effelsberg.
This paper includes archived data obtained through the~Australia Telescope Online Archive and the
CSIRO Data Access Portal (http://datanet.csiro.au/dap/).
We also thank M.~Dembska, K.~Lazaridis and K.~Maciesiak for technical assistance. 
JK, LW, and OM acknowledge the support of the Polish Grant N N203 391934. 
\end{acknowledgements}

\end{document}